\documentstyle[psfig]{mn}
\newcommand\etal{et al. }
\newcommand\refer{\par \noindent\hangindent=3pc \hangafter=1}

\newcommand\ros{{\it ROSAT}}
\newcommand\asca{{\it ASCA}}

\newcommand\eg{eg }

\newcommand\ie{i.e. }
\newcommand\ecs{erg cm$^{-2}$ s$^{-1}$}
\newcommand\nh{N_{H}}

\newcommand\halpha{H$\alpha$}

\title{A Survey of hard spectrum \ros\ sources 2: optical identification of hard
sources}
\author[Page, Mittaz \& Carrera]{M.J. Page\(^{1}\), J.P.D. Mittaz\(^{1}\), 
F.J. Carrera\(^{1,2}\)\\
\(^{1}\)Mullard Space Science Laboratory, University College London,
Holmbury St Mary, Dorking, Surrey RH5 6NT, UK.\\
\(^{2}\)Instituto de F\'\i sica de Cantabria (Consejo Superior de
Investigaciones Cient\'\i ficas--Universidad de Cantabria), 39005
Santander, Spain.}

\date{}

\begin{document}
\maketitle

\begin{abstract}
 
We have surveyed 188 \ros\ PSPC fields for X--ray sources with hard spectra
(\(\alpha < 0.5\)); such sources must be major contributors to the X--ray
background at faint fluxes.  In this paper we present optical identifications
for 62 of these sources: 28 AGN which show broad lines in their optical spectra
(BLAGN), 13 narrow emission line galaxies (NELGs), 5 galaxies with no visible
emission lines, 8 clusters and 8 Galactic stars.

The BLAGN, NELGs and galaxies have similar distributions of X--ray flux and
spectra.  Their \ros\ spectra are consistent with their being AGN obscured by
columns of \(20.5 < {\rm log} (\nh/{\rm cm^{-2}}) < 23\).  
The hard spectrum BLAGN have a distribution of X--ray to optical ratios which
is similar to that found for AGN from soft X--ray surveys (\(1 <
\alpha_{OX} < 2\)).  However, a relatively large proportion (15\%) of the
BLAGN, NELGs and galaxies are radio loud. This could be because the radio jets
in these objects produce intrinsically hard X--ray emission, or if their
hardness is due to absorption, it could be because radio loud objects are more
X--ray luminous than radio quiet objects.
The 8 hard sources identified as clusters of galaxies are the brightest, and
softest group of sources and hence clusters are unlikely to be an important
component of the hard, faint population.

We propose that BLAGN are likely to constitute a significant fraction of the
faint, hard, 0.5 - 2 keV population and could be important to reproducing the
shape of the X--ray background, because they are the most numerous type of
object in our sample (comprising almost half the identified sources), and
because all our high redshift (\(z>1\)) identified hard sources have broad
lines.

\end{abstract}

\section{Introduction}

The origin of most of the X--ray emission in the Universe is still unknown
because the sources that produce most of the \(>\) 2 keV X--ray background
(XRB) are still to be resolved.  \ros\ surveys have succeeded in resolving
\(\sim 80\%\) of the 1-2 keV XRB into individual sources (Hasinger \etal 1998),
and optical identification and X--ray spectroscopy has been possible for
brighter sources which produce \(\sim 40\%\) of the 1-2 keV background. The
majority of these sources are broad line AGN (hereafter BLAGN), and at faint
fluxes narrow emission line galaxies (hereafter NELGs, McHardy \etal 1998);
Schmidt \etal (1998) argued that the NELGs are also AGN, but with low
luminosity or obscured broad line regions.  On average, faint NELGs have harder
X--ray spectra (\( f_{\nu} \propto \nu^{-\alpha}\) with \(\alpha \sim 0.5\),
 Romero-Colmenero \etal 1996, Almaini \etal
1996) than the broad line AGN which have mean \(\alpha \sim 1\) (Mittaz \etal
1999, Ciliegi \etal 1994).
  
Despite the success of \ros\ surveys, the XRB cannot be synthesised by
extrapolating the observed source populations to faint fluxes, because the
resultant spectrum would be softer than that of the background; 
this discrepancy is present for all energy bands between 0.5 and 40 keV.
This means that at faint fluxes there must be a population of sources with
spectra that are harder than the background.  According to leading models,
these hard sources (\eg Fabian 1999, Gilli, Risaliti \& Salvati 1999) are
obscured AGN. However, the physical nature and observational appearance of the
XRB producing population is not yet known, and is the subject of some
debate. For example, Gilli, Risaliti \& Salvati (1999) examined a model
intended to reproduce the XRB by extrapolating the X--ray emission of present
epoch AGN to high redshift using the observed soft X--ray luminosity
function. In contrast, the model of Fabian (1999) has a large fraction of the
XRB due to a population of high redshift, heavily obscured, growing AGN, which
are different to anything observed in the local universe.

In Page \etal (2000), hereafter paper 1, we presented a catalogue of
147 serendipitous \ros\ sources which have spectra harder than that of the XRB.
These sources have a steep \(N(S)\) relation down to the sensitivity limit of
our survey (\(\sim 10^{-14}\) \ecs ), and are therefore likely to be the bright
tail of the population of hard sources that dominate the source counts at faint
fluxes. As such, they could offer us a preview of the faint, hard and dominant,
X-ray source population. 
We have therefore undertaken a programme of optical and infrared observations
of our \ros\ hard source sample to find out what these sources are.
In this paper we present the hard source identifications
obtained from our optical (spectroscopic) campaign together with those 
found in existing  
catalogues. Section \ref{sec:method} details our method and observations, the
results of which are given in Section \ref{sec:results}. These results
are discussed in the context of absorbed AGN and the XRB in Section
\ref{sec:discussion}. Finally, we 
present our conclusions in Section \ref{sec:conclusions}.

Throughout this paper we define power law spectral index $\alpha$ such that 
\(f_{\nu} \propto \nu^{-\alpha}\).

\section{Optical identification}
\label{sec:method}
\subsection{Strategy}

To produce a systematic and efficient optical identification
programme, we divided the hard X-ray sources into two groups depending on the
ease of optical identification. Sources which had one or two plausible 
candidates present in APM data and/or the DSS were considered suitable for
optical spectroscopy and make up the `spectroscopic sample', while sources with
no plausible candidates, or more than two, were considered unsuitable for
spectroscopy and became the `imaging sample'. There are 103 sources in the
spectroscopic sample and 44 sources in the imaging sample.
This paper will deal only with
identified sources from the spectroscopic sample (except for two sources,
RXJ005812.20-274217.8 and RXJ101112.05+554451.3, which are fainter than our
spectroscopic sample limit but have catalogue identifications). 
The properties of sources in the imaging sample, and their 
relationship to the spectroscopic sample sources, will be discussed briefly in 
Section \ref{sec:imaging} and in more detail in Carrera
\etal (in preparation) where we will present the results of our optical and infrared
imaging. The three sources of data for the identification process were
spectra taken on the William Herschel Telescope (WHT) and the European Southern
Observatory 3.6m Telescope (ESO 3.6), and databases of existing catalogues.
The optical spectra will be presented in Mittaz \etal (2001) 
along with a full description of the observations and data 
reduction.

Catalogue identifications were obtained by searching the NASA Extragalactic
Database (NED) and SIMBAD around all the \ros\ hard source positions (not just
the spectroscopic sample). One further source (RXJ043420.48-082136.7) was
identified from a WHT ISIS spectrum taken during the RIXOS programme, 
but was not part of the
final RIXOS sample presented in Mason \etal (2000).

\section{Results}
\label{sec:results}

 \begin{table*}
 \tabcolsep=1mm
 \caption{Hard source optical counterparts}
 \label{tab:IDs}
 \begin{tabular}{lccllcll}
 &&&&&&&\\
                                                          Source      & \multicolumn{2}{c}{optical position}& origin & type & $z$ & cat name & notes\\
                                                                  &RA&dec&&&&&\\
RXJ001144.43-362638.0& 00 11 44.54 & -36 26 39.0 &EFOSC  &BLAGN   & 0.900&                              &                              \\
RXJ004651.98-204329.0& 00 46 51.83 & -20 43 28.6 &cat    &BLAGN   & 0.380&[HB89] 0044-209               &                              \\
RXJ005734.78-272827.4& 00 57 34.94 & -27 28 28.0 &EFOSC  &BLAGN   & 2.185&                              &                              \\
RXJ005736.81-273305.9& 00 57 36.75 & -27 33 04.6 &cat    &NELG    & 0.213&GSGP 4X:069                   &                              \\
RXJ005746.75-273000.8& 00 57 46.83 & -27 30 00.9 &cat    &galaxy  & 0.019&ESO 411- G 034                &                              \\
RXJ005801.64-275308.6& 00 58 01.32 & -27 53 10.2 &EFOSC  &NELG    & 0.416&GSGP 4X:091                   &                              \\
RXJ005812.20-274217.8& 00 58 13.57 & -27 42 11.4 &cat    &NELG    & 0.597&GSGP 4X:100                   &                              \\
RXJ013555.47-183210.2& 01 35 55.71 & -18 32 24.9 &EFOSC  &star    & 0.000&                              &                              \\
RXJ013707.63-183846.4& 01 37 07.75 & -18 38 49.9 &EFOSC  &star    & 0.000&                              &Unlikely to be X-ray source   \\
RXJ013721.47-182558.3& 01 37 21.94 & -18 26 03.0 &EFOSC  &NELG    & 0.336&                              &                              \\
RXJ014159.22-543037.0& 01 41 59.81 & -54 30 39.6 &EFOSC  &BLAGN   & 0.168&                              &                              \\
RXJ031456.58-552006.8& 03 14 56.30 & -55 20 06.1 &cat    &galaxy  & 0.387&[GZd97] 1.4GHz 38             &                              \\
RXJ033340.22-391833.4& 03 33 39.54 & -39 18 41.4 &EFOSC  &BLAGN   & 1.436&                              &                              \\
RXJ033402.54-390048.7& 03 34 03.26 & -39 00 36.9 &EFOSC  &galaxy  & 0.061&PKS 0332-39                   &WAT radio source              \\
RXJ034119.02-441033.3& 03 41 19.23 & -44 10 29.9 &cat    &BLAGN   & 0.505&QSF3X:51                      &                              \\
RXJ043420.48-082136.7& 04 34 20.19 & -08 21 31.3 &cat    &BLAGN   & 0.155&                              &                              \\
RXJ045558.99-753229.1& 04 55 58.82 & -75 32 28.0 &cat    &NELG    & 0.018&ESO 033- G 002                &                              \\
RXJ052839.93-325148.5& 05 28 39.83 & -32 51 44.7 &cat    &cluster & 0.273&[VMF98] 042                   &                              \\
RXJ082640.20+263112.3& 08 26 40.52 & +26 31 14.1 &ISIS   &NELG    & 0.182&                              &                              \\
RXJ085340.52+134924.9& 08 53 41.01 & +13 49 19.7 &cat    &NELG    & 0.190&MS 0850.8+1401                &Uncertain ID                  \\
RXJ085851.49+141150.7& 08 58 50.75 & +14 11 54.5 &ISIS   &NELG    & 0.453&                              &                              \\
RXJ090518.27+335006.0& 09 05 17.94 & -33 50 16.1 &ISIS   &NELG    & 0.425&                              &                              \\
RXJ090923.64+423629.2& 09 09 23.82 & +42 36 23.3 &ISIS   &BLAGN   & 0.177&                              &                              \\
RXJ091908.27+745305.6& 09 19 10.56 & +74 53 11.2 &ISIS   &galaxy  & 0.073&                              &                              \\
RXJ094144.51+385434.8& 09 41 44.61 & +38 54 39.1 &ISIS   &BLAGN   & 1.819&                              &                              \\
RXJ095340.67+074426.1& 09 53 40.12 & +07 44 12.3 &cat    &BLAGN   & 0.760&RIXOS F218\_021               &Uncertain ID                  \\
RXJ101008.53+513334.9& 10 10 08.91 & +51 33 31.0 &ISIS   &star    & 0.000&                              &                              \\
RXJ101112.05+554451.3& 10 11 12.30 & +55 44 47.0 &cat    &BLAGN   & 1.246&87GB 100755.1+560014          &                              \\
RXJ101123.17+524912.4& 10 11 22.67 & +52 49 12.3 &ISIS   &BLAGN   & 1.012&                              &                              \\
RXJ101147.48+505002.2& 10 11 47.82 & +50 49 57.5 &ISIS   &BLAGN   & 0.079&5H 23                         &                              \\
RXJ104648.27+541235.4& 10 46 48.88 & +54 12 19.4 &ISIS   &star    & 0.000&                              &                              \\
RXJ104723.37+540412.6& 10 47 23.50 & +54 04 06.7 &ISIS   &BLAGN   & 1.500&                              &                              \\
RXJ110742.05+723236.0& 11 07 41.59 & +72 32 35.8 &cat    &BLAGN   & 2.100&[HB89] 1104+728               &                              \\
RXJ111750.51+075712.8& 11 17 50.78 & +07 57 11.4 &cat    &BLAGN   & 0.698&RIXOS F258\_001               &                              \\
RXJ111926.34+210646.1& 11 19 25.93 & +21 06 47.4 &cat    &cluster & 0.176&                              &                              \\
RXJ111942.16+211518.1& 11 19 42.13 & +21 15 16.6 &cat    &BLAGN   & 1.288&                              &                              \\
RXJ112056.87+132726.2& 11 20 57.48 & +13 27 08.0 &ISIS   &star    & 0.000&                              & Unlikely to be X-ray source  \\
RXJ114621.27+285320.6& 11 46 19.93 & +28 53 06.6 &cat    &cluster & 0.170&part of [VMF98] 107           &                              \\
RXJ115952.10+553212.1& 11 59 52.28 & +55 32 06.1 &cat    &cluster & 0.081&MS 1157.3+5548                &                              \\
RXJ120403.79+280711.2& 12 04 03.55 & +28 07 01.3 &cat    &cluster & 0.167&MS 1201.5+2824                &                              \\
RXJ121017.25+391822.6& 12 10 16.61 & +39 18 16.6 &ISIS   &NELG    & 0.022&                              &                              \\
RXJ121115.30+391146.8& 12 11 15.67 & +39 11 54.2 &cat    &cluster & 0.340&MS 1208.7+3928                &                              \\
RXJ121803.82+470854.6& 12 18 4.54  & +47 08 51.0 &ISIS   &BLAGN   & 1.743&                              &                              \\
RXJ124913.86-055906.2& 12 49 13.85 & -05 59 19.4 &cat    &BLAGN   & 2.212&[HB89] 1246-057               &BALQSO                        \\
RXJ131635.62+285942.7& 13 16 34.72 & +28 59 29.3 &cat    &BLAGN   & 0.277&RIXOS F224\_026               &                              \\
RXJ133146.37+111420.4& 13 31 45.93 & +11 14 12.8 &ISIS   &star    & 0.000&                              & Unlikely to be X-ray source  \\
RXJ133147.00+105653.0& 13 31 46.58 & +10 56 55.7 &cat    &star    & 0.000&RIXOS F278\_026               &                              \\
RXJ133152.51+111643.5& 13 31 52.25 & +11 16 49.6 &cat    &BLAGN   & 0.090&RIXOS F278\_010               &                              \\
RXJ135105.69+601538.5& 13 51 06.33 & +60 15 39.5 &ISIS   &NELG    & 0.291&                              &                              \\
RXJ135529.59+182413.6& 13 55 29.54 & +18 24 21.3 &cat    &BLAGN   & 1.196&RIXOS F268\_011               &                              \\
RXJ140134.94+542029.2& 14 01 34.59 & +54 20 31.1 &ISIS   &galaxy  & 0.069&                              &                              \\
RXJ140416.61+541618.2& 14 04 16.79 & +54 16 14.6 &ISIS   &BLAGN   & 1.405&                              &                              \\
RXJ142754.71+330007.0& 14 27 54.51 & +32 59 59.8 &cat    &BLAGN   & 0.420&RIXOS F110\_034               &Uncertain ID                  \\
RXJ163054.25+781105.1& 16 30 55.00 & +78 11 03.9 &cat    &BLAGN   & 0.358&                              &                              \\
RXJ163308.57+570258.7& 16 33 08.59 & +57 02 54.8 &ISIS   &BLAGN   & 2.802&WN B1632+5709                 &                              \\
RXJ170041.60+641259.0& 17 00 41.71 & +64 12 58.4 &ISIS   &cluster & 0.230&ABELL 2246                    &                              \\
RXJ170123.32+641413.0& 17 01 23.47 & +64 14 11.8 &cat    &cluster & 0.440&[RTH97] B                     &                              \\
RXJ204640.48-363147.5& 20 46 40.13 & -36 31 48.1 &EFOSC  &BLAGN   & 1.122&                              &                              \\
RXJ204716.74-364715.1& 20 47 16.75 & -36 47 23.2 &EFOSC  &NELG    & 0.050&                              &                              \\
RXJ213807.61-423614.3& 21 38 07.97 & -42 36 17.8 &EFOSC  &BLAGN   & 0.019&ESO 287- G 042                &                              \\
RXJ223619.89-261426.2& 22 36 20.43 & -26 14 37.4 &EFOSC  &star    & 0.000&                              &                              \\
RXJ235113.89+201347.3& 23 51 13.91 & +20 13 46.4 &cat    &NELG    & 0.043&MCG +03-60-031                &                              \\
 \end{tabular}
 \end{table*}

Table \ref{tab:IDs} contains the list of optical identifications.  
We have categorised our sources as follows.
X-ray sources with an
optical counterpart which is a Galactic star, of any type, have been classified
as stars, and X-ray sources which are clusters of galaxies have been
classified as clusters.  
X-ray sources with an extragalactic optical
counterpart which has one or more broad emission lines, or a broad component
(\(>2000\) km s\(^{-1}\)) to an emission line, is classified as a broad line
AGN (BLAGN).  X-ray sources with galaxy optical counterparts showing narrow
emission lines (\(<2000\) km s\(^{-1}\)) but no broad emission lines are
classed as narrow emission line galaxies (NELGs).  Note that none of the narrow
line objects in this survey would be classified as  
narrow line Seyfert 1s (\eg by the criteria given in Goodrich 1989). 
Galaxies with no
discernable emission lines were classed as galaxies. 
The numbers of objects in each category, and their mean properties are given in
Table \ref{tab:numbers}.
The sample is dominated by
extragalactic sources, and of these over half are broad line AGN.

\begin{figure}
\begin{center}
\leavevmode \psfig{figure=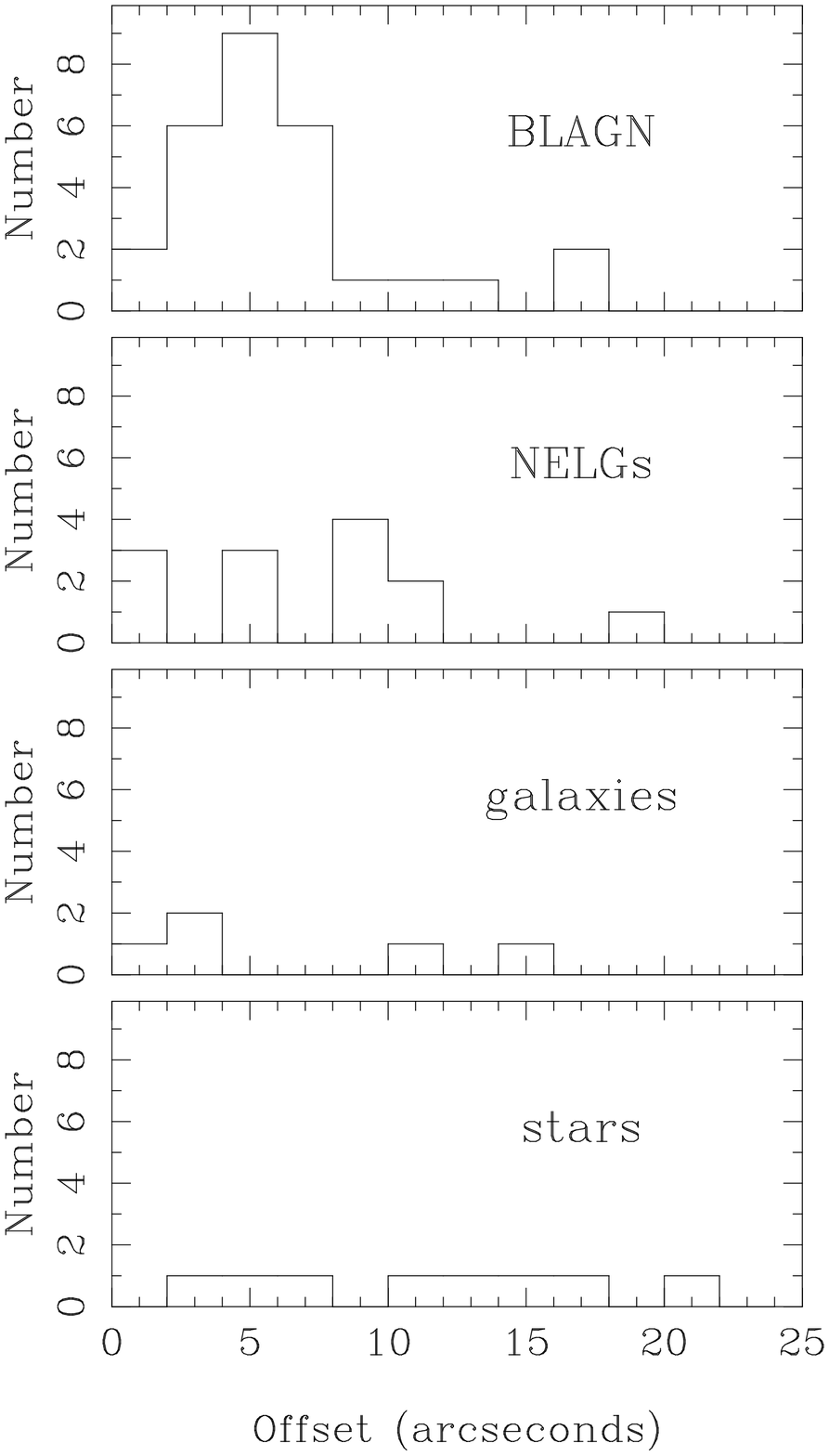,width=80truemm}
\caption{Histogram of offsets, in bins of 2 arcseconds,
 between the X--ray positions and optical
counterparts for the different classes of identified sources.}
\label{fig:offsets}
\end{center}
\end{figure}

The distributions of offsets between the X-ray positions and the optical
counterparts for the different classes of hard source are shown in Fig. 
\ref{fig:offsets}. The mean offsets for each source type are given in Table
\ref{tab:numbers} and are 6 -- 7 arcseconds for all source types except for the
Galactic stars, which have a mean offset of 12 arcseconds and a flat
distribution of offsets (Fig. \ref{fig:offsets}); this suggests that the `star'
identifications are less secure than the others.

We have systematically checked the redshifts of the extragalactic hard sources
for similarity to the redshifts of the targets of the PSPC observations in
which they were detected. Only two sources, RXJ111926.34+210646.1 and
RXJ120403.79+280711.2, both clusters of galaxies, have redshifts similar to the
observation targets.

\begin{table}
\caption{Different types of identified hard sources and their mean properties.
N is the number of sources, $\langle S \rangle$ is the mean fitted flux in 
units of
$10^{-14}$ \ecs, $\langle \alpha \rangle$ is the mean fitted spectral slope, 
$\langle z \rangle$ is the mean
redshift and $\langle \log \ L \rangle$ is the mean of the log of 
observed 0.5 - 2 keV luminosity in
erg s$^{-1}$, calculated as described in Section \ref{sec:redlums}. For
comparison, the 
unidentified sources from the spectroscopic sample are listed as 
No ID (S), while the
unidentified sources from the imaging sample are listed as `No ID (I)'.}
\label{tab:numbers}
 \begin{tabular}{lcccccc}
 &N&$\langle S \rangle$&$\langle{\alpha}\rangle$&$\langle z \rangle$&
 $\langle \log\  L\rangle$&$\langle offset \rangle$\\
     BLAGN&  28&   8.1& -0.29&  1.00& 43.71&   6.1\\
     NELGs&  13&  13.6& -0.74&  0.25& 42.68&   7.2\\
  galaxies&   5&   4.0& -0.43&  0.12& 41.85&   6.5\\
  clusters&   8&  19.9&  0.21&  0.23& 43.38& - \\
     stars&   8&   3.1& -0.91& - & - & 11.45\\
 No ID (S)&  43&   3.9& -0.52& - & - & - \\
 No ID (I)&  42&   3.2& -0.54& - & - & - \\
 \end{tabular}
\end{table}

\subsection{X-ray slopes and fluxes}
\label{sec:slopeflux}

\begin{figure*}
\begin{center}
\leavevmode \psfig{figure=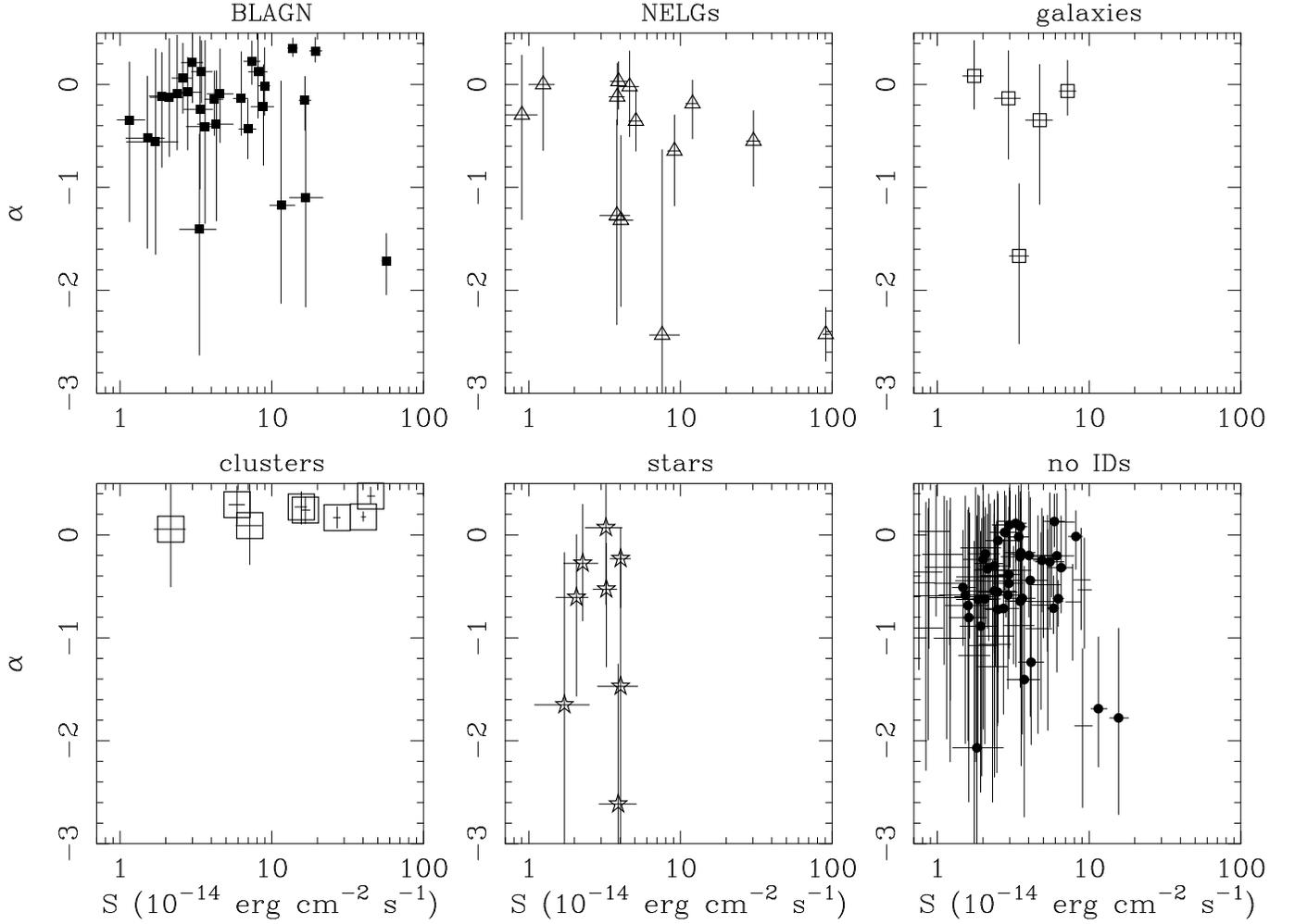,width=180truemm,angle=270}
\caption{X-ray spectral slopes and fluxes of the different 
types of hard source. For comparison the unidentified sources are shown in the
last panel: the spectroscopic sample as filled
dots and the imaging sample as crosses (see Section \ref{sec:method} for
definitions of the two samples).}
\label{fig:specflux}
\end{center}
\end{figure*}

The fitted X-ray spectral slopes and fluxes of the different types of sources 
are
shown in Fig. \ref{fig:specflux}, and their mean slopes and fluxes are given in
Table \ref{tab:numbers}.  A noticeable feature of
Fig. \ref{fig:specflux} is that the clusters of galaxies are concentrated
towards the top right (soft spectrum, high flux) corner compared to the other
sources. This trend is confirmed by a two dimensional Kolmogorov Smirnov (2DKS)
test (Fasano \& Franceschini 1987): the cluster distribution of (\(\alpha, S\))
is different to that for any (or all) of the other source types with \(>99\%\)
confidence.  Because they are not particularly hard, and because they are
predominantly bright and therefore do not have as steep an \(N(S)\) relation as
the rest of the hard sources, the clusters are unlikely to be important
contributors to the faint hard source population.  

Note that the very hard mean spectral index of the Galactic stars (see Table
\ref{tab:numbers}) is due to the three objects with best fit 
\(\alpha < -1\), two of 
which (RXJ112056.87 +132726.2 with best fit \(\alpha = -1.65\) and 
RXJ133146.37+111420.4 with best fit \(\alpha = -1.472\)) are unlikely to be
correct identifications (see Section \ref{sec:notes}).

The other three types of sources 
(BLAGN, NELGs and galaxies) have distributions of
(\(\alpha, S\)) which are indistinguishable from one another according to the
2DKS test; any (or all) could be significant contributors to the faint, hard
population.

For comparison Figure \ref{fig:specflux} also shows the unidentified sources.
Unidentified objects in the spectroscopic sample are found with similar
spectral slopes and fluxes to the identified sources (except the clusters which
are softer and brighter). The sources in the imaging sample are more
concentrated toward faint fluxes and comprise most of the faintest (\(S \sim
10^{-14}\) \ecs) hard sources.

\subsection{The unidentified sources}
\label{sec:imaging}

We now consider briefly how fair a subsample the identified sources
are of the whole \ros\ hard source sample. Starting with the
spectroscopic sample, candidates were chosen for identification on our
WHT and ESO 3.6m observing runs without regard to their optical
magnitudes and morphologies, (except to exclude sources with an
existing catalogue identification) and are hence an unbiased subsample
of the spectroscopic sample. The catalogue identifications, although
including a few optically bright galaxies, are mostly from previous
flux limited X--ray surveys. They therefore tend to have higher X--ray
fluxes than the spectroscopic sample as a whole, but are otherwise a
fair subsample. This means that except for a bias towards higher
X--ray fluxes, the identified sources are a fair subsample of the
spectroscopic sample.

\begin{figure}
\begin{center}
\leavevmode
\psfig{figure=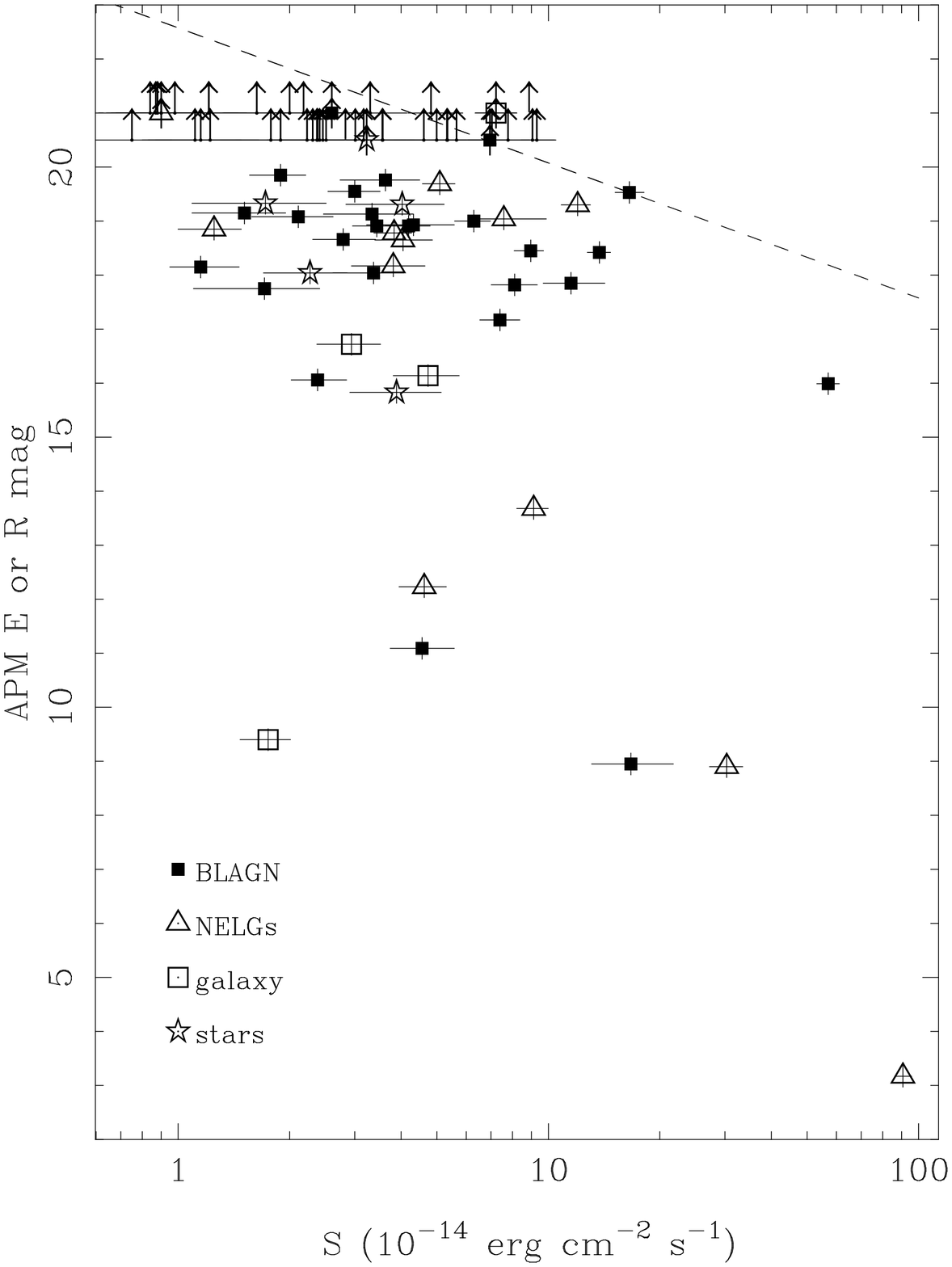,width=80truemm}
\caption{Optical magnitudes against X--ray fluxes for the identified sources inthe spectroscopic sample (symbols) and the sources in the imaging sample (lower limits). The gradient of the dashed line corresponds to a constant ratio of 
X--ray to optical flux.}
\label{fig:optx}
\end{center}
\end{figure}

The spectroscopic sample is a significant fraction of the whole \ros\
hard source sample (103 of 147 sources), but it cannot be considered a
fair subsample because it excludes the optically faint sources which
form the imaging sample.  Without spectroscopically identifying the
imaging sample (which is currently impractical) it is not possible to
determine whether or not the imaging sample is {\it actually} made up
of the same mix of sources as the spectroscopic sample. However we can
examine whether it is {\it possible}, by assuming that similar types
of sources will have similar X--ray to optical flux ratios. Figure
\ref{fig:optx} shows the X--ray flux against the APM red (E or R)
magnitude for the identified sources with available APM data. BLAGN,
NELGs and galaxies are found with similar ranges of X--ray to optical
ratio. The optical counterparts to the imaging sample sources are
presumed to be fainter than the plate limit, which we take to be 20.5
and 21.0 for sources with APM data available for E and R plates
respectively.  The imaging sample sources are plotted in
Fig. \ref{fig:optx} with these lower limits except for two sources 
RXJ031956.76-663938.5
and RXJ101031.05+503458.6, which are in the imaging sample because 
there are too many
potential counterparts for practical spectroscopic follow up.  The
dashed line corresponds to an X--ray to optical flux ratio which is
similar to that of RXJ111750.51+075712.8, which has the highest X--ray to 
optical flux
ratio of the identified spectroscopic sample sources excluding the
clusters. Only six of the imaging sample sources lie above the dashed
line, and therefore definitely have X--ray to optical flux ratios
different to those of the identified sources.  Therefore the limits on
the optical to X--ray ratios are consistent with
the identified sources being the same types of objects as constitute
almost the whole \ros\ hard source sample.

\subsection{Redshifts and luminosities}
\label{sec:redlums}

\begin{figure}
\begin{center}
\leavevmode
\psfig{figure=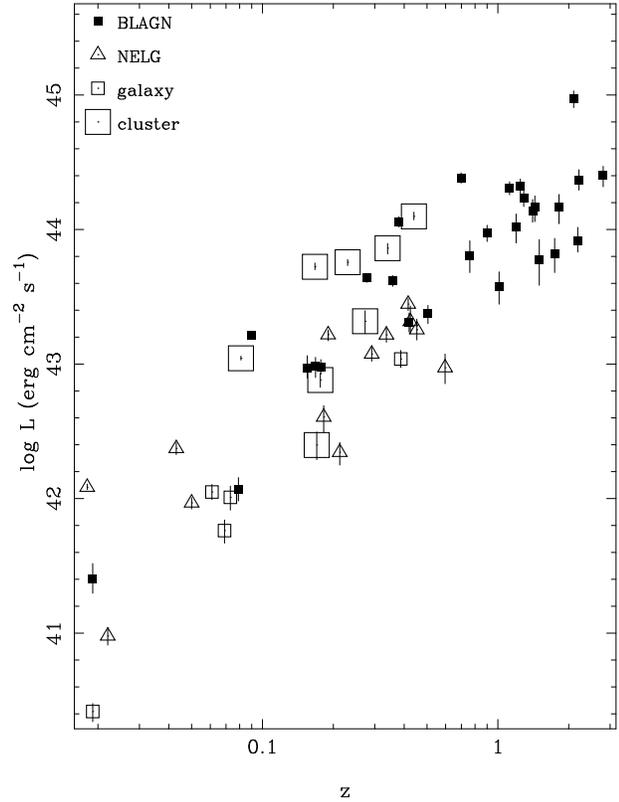,width=80truemm}
\caption{Redshifts and observed 0.5 - 2 keV luminosities for the different extragalactic 
source types}
\label{fig:zlumi}
\end{center}
\end{figure}

Fig. \ref{fig:zlumi} shows the redshifts and 0.5 - 2 keV luminosities for the
extragalactic hard sources. Luminosities were calculated from the fitted
fluxes, K corrected using the best fit spectral slopes, and assuming \(H_{0} =
50\) km s\(^{-1}\) Mpc\(^{-3}\), \(q_{0} = 0\). The majority of the BLAGN have
higher luminosities and redshifts than the NELGs and galaxies; all the high
redshift (\(z>1\)) sources are BLAGN. This may imply that either hard 
spectrum narrow line sources
do not exist at high luminosity/redshift or our spectroscopic sample is not 
deep enough (in X--ray or optical flux) to find them.

\subsection{X--ray - optical - radio flux ratios}
\label{sec:alphaoxr}

It is common to parameterise the X--ray to optical and optical to radio flux
ratios of AGN as \(\alpha_{OX}\) and \(\alpha_{OR}\), where \(\alpha_{OX}\) is
the slope of the power law which (in the object's rest frame) would connect the
flux density (F\(_{\nu}\)) at 2500 \AA\ with the X--ray flux density at 2 keV,
and \(\alpha_{OR}\) is the slope of the power law which would connect the flux
density at 2500 \AA\ with the flux density at 5 GHz.

For all sources we have estimated the rest frame 2500 \AA\ flux density using B
photometry (taken from Carrera \etal in preparation) and assuming a power
law optical - UV spectrum with slope \(\alpha=0.5\), and using the B magnitude
to flux conversion given by Wilkes \etal (1994). The B magnitudes have been
corrected for Galactic reddening using the expression for \(E(B-V)\) in Bohlin,
Savage \& Drake (1978) and assuming \(A_{B}=4E(B_V)\). 

To estimate the rest frame 5 GHz fluxes we have searched for potential radio
counterparts to our X--ray sources in the catalogues of
radio sources in the VLA 1.4 GHz FIRST (White \etal 1997) and NVSS (Condon
\etal 1998) surveys and the 5 GHz
Parkes-MIT-NRAO (PMN, Griffith \& Wright 1993) survey. Additionally, radio
images from these surveys were searched by eye to ensure that any extended 
radio sources associated with our X--ray sources were not missed, and to 
ensure that there were no spurious radio counterparts associated with other 
extended radio sources. The uncertainty in radio source position from these
three surveys ranges from $\sim$ better than 1 arcsecond (FIRST) to around 10
arcseconds (PMN). The sky density of radio sources is sufficiently low
(reaching around 100 deg$^{-2}$ at the $\sim$1 mJy completeness limit of FIRST)
that it is unlikely that there are any chance coincidences. Where no suitable
radio source appears in a survey catalogue, we have taken the catalogue
completeness limit as the upper limit to the radio flux of the X--ray
source. We took the best measurement (or upper limit) available from the 
three surveys and assume a radio spectral slope of \(\alpha =
0.7\). 

The 2 keV flux density has been estimated from the fitted 0.5 - 2 keV fluxes
and assuming an X--ray spectral slope of \(\alpha=0\). 

\begin{figure}
\begin{center}
\leavevmode
\psfig{figure=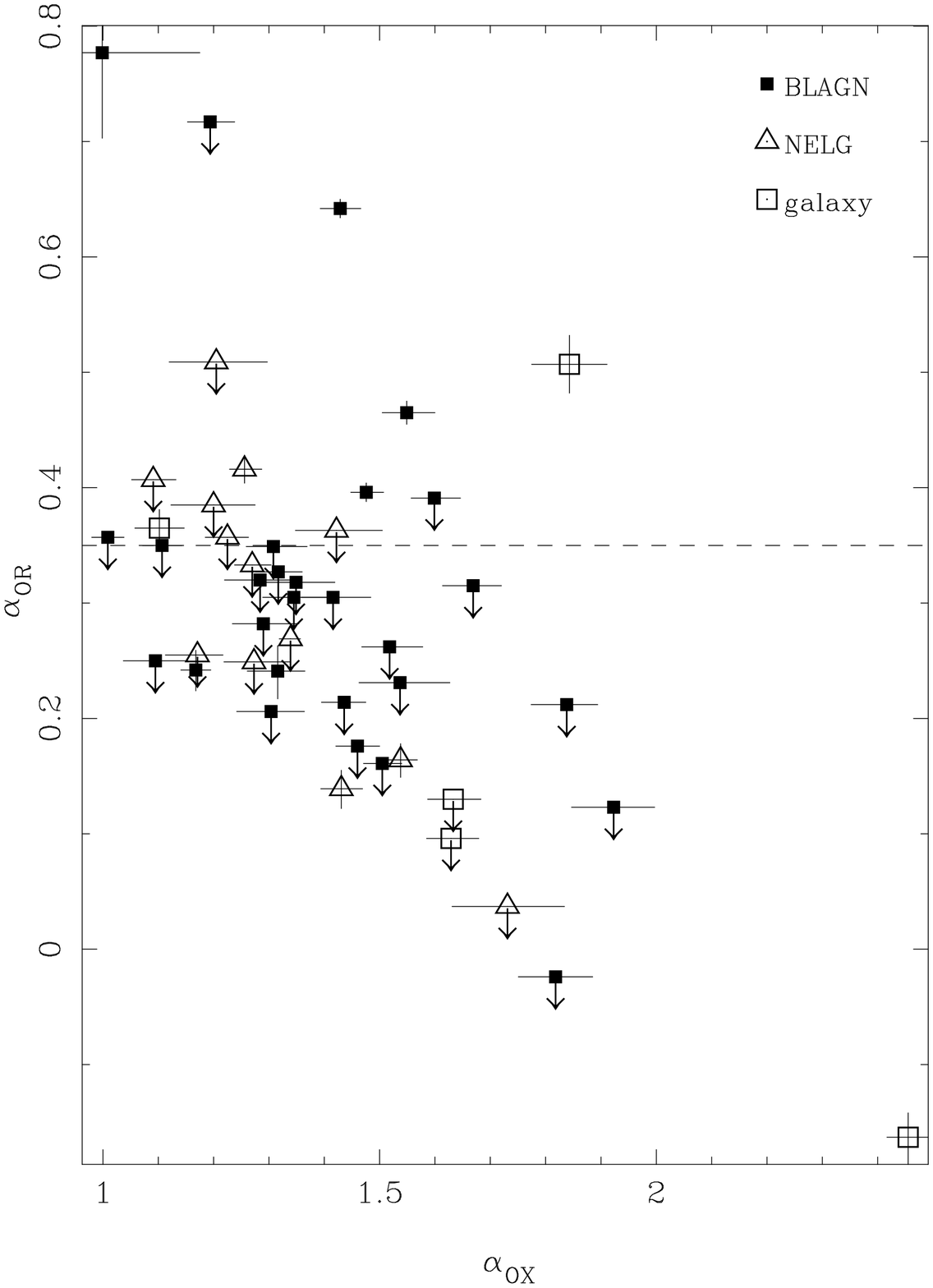,width=80truemm}
\caption{$\alpha_{OX}$ and $\alpha_{OR}$ for the hard sources}
\label{fig:alphaoxr}
\end{center}
\end{figure}

We then computed 
\[
\alpha_{OX} = 0.384\ {\rm log} [f_{\nu}(2500{\rm
\AA})/f_{\nu}(2 {\rm keV})]
\]
and 
\[
\alpha_{OR} = -0.186\ {\rm log} [f_{\nu}(2500{\rm
\AA})/f_{\nu}(5 {\rm GHz})]
\]

\(\alpha_{OX}\) and \(\alpha_{OR}\) are shown in Fig. \ref{fig:alphaoxr} for
all the hard sources identified as BLAGN, NELGs and galaxies. The horizontal
dashed line marks \(\alpha_{OR}=0.35\) which is commonly used to differentiate
radio loud and radio quiet objects (Zamorani \etal 1981). Seven hard sources
(four BLAGN, 2 galaxies and 1 NELG) lie above this line, and are therefore
radio loud. This is a large radio loud fraction (15$^{+7}_{-5}$\% where errors
are the Poission 68\% confidence interval, Gehrels 1986) 
of the 
BLAGN, NELGs and
galaxies compared to the fraction found in normal \ros\ surveys {\it without}
spectral selection, \eg Cilliegi \etal (1995) find that only two of the eighty
CRSS AGN and NELGs (2.5$^{+3.3}_{-1.6}$\%) are radio loud.  
On the other hand, the distribution
of \(\alpha_{OX}\) of the hard sources is quite similar to that found in other
AGN surveys (\eg Ciliegi \etal 1995, Wilkes \etal 1994), with the majority of
sources having \(1<\alpha_{OX}<1.8\).

\subsection{X--ray absorption}

The leading hypothesis to explain the majority of faint hard sources which
contribute substantially to the XRB is that they are intrinsically absorbed AGN
(\eg Fabian \& Iwasawa 1999, Setti \& Woltjer 1989). This is also a likely
hypothesis for our hard sources, because the optical spectra and colours of a
significant fraction of our BLAGN and NELGs suggest absorption (Mittaz \etal
2001 and Carrera \etal 2001).  Assuming that our BLAGN, NELGs and galaxies have
hard spectra because of absorption, we have estimated their column densities
from their 3 colour \ros\ PSPC spectra. We assume that before absorption they
have power law X--ray spectra with energy index \(\alpha=1\) (typical for X-ray
selected AGN, \eg Mittaz \etal 1999, Maccacaro \etal 1988) and fit their
intrinsic cold gas column. The fitting procedure was identical to that
described in paper 1 except that the free parameters in
the fit are intrinsic column and power law normalisation rather than power law
slope and normalisation. The results of these fits are given in Table
\ref{tab:oxrcolumns}.

 \begin{table*}
 \tabcolsep=1mm
 \caption{Hard source $\alpha_{OX}$, $\alpha_{OR}$ and fitted columns}
 \label{tab:oxrcolumns}
 \scriptsize
 \begin{tabular}{lcccccccccccc}
         Source      &\multicolumn{3}{c}{--------- X--ray ---------}& B mag & $A_{B}$ & \multicolumn{3}{c}{------------ radio ------------} & $\alpha_{OX}$ &$\alpha_{OR}$ & \multicolumn{2}{c}{--- X--ray column fit ---}\\
                                                                  & flux$^{a}$ & $\alpha$ & log $L^{b}$ & & & flux & $\nu$ & survey &  & & $\nh$ & norm$^{c}$ \\
 &&&&&&(mJy)&(GHz)&&&&(cm$^{-2}$)&\\
 &&&&&&&&&&&&\\
RXJ001144.43-362638.0&   4.19$^{  +0.60}_{  -0.56}$&  -0.14$^{  +0.27}_{  -0.35}$&  43.97$^{  +0.06}_{  -0.06}$&  20.38$\pm$    0.12&   0.08&$<$   2.50&    1.4&NVSS      &   1.32$^{  +0.04}_{  -0.04}$&$<$   0.33&  21.34$^{  +0.43}_{  -0.29}$&   2.32$^{  +0.52}_{  -0.39}$\\
RXJ004651.98-204329.0&  19.37$^{  +1.95}_{  -1.56}$&   0.32$^{  +0.13}_{  -0.11}$&  44.05$^{  +0.04}_{  -0.04}$&  18.60$\pm$    0.30&   0.11&$<$   2.50&    1.4&NVSS      &   1.30$^{  +0.06}_{  -0.06}$&$<$   0.21&  20.72$^{  +0.08}_{  -0.09}$&   9.91$^{  +0.89}_{  -0.92}$\\
RXJ005734.78-272827.4&   1.15$^{  +0.31}_{  -0.20}$&  -0.34$^{  +0.56}_{  -0.99}$&  43.91$^{  +0.10}_{  -0.08}$&  18.62$\pm$    0.16&   0.13&$<$   2.50&    1.4&NVSS      &   1.84$^{  +0.06}_{  -0.06}$&$<$   0.21&  22.60$^{  +0.26}_{  -0.48}$&   0.90$^{  +0.33}_{  -0.27}$\\
RXJ005736.81-273305.9&   1.25$^{  +0.23}_{  -0.25}$&   0.00$^{  +0.36}_{  -0.64}$&  42.34$^{  +0.07}_{  -0.10}$&  20.72$\pm$    0.30&   0.13&$<$   2.50&    1.4&NVSS      &   1.42$^{  +0.08}_{  -0.07}$&$<$   0.36&  21.07$^{  +0.40}_{  -0.46}$&   0.79$^{  +0.22}_{  -0.18}$\\
RXJ005746.75-273000.8&   1.75$^{  +0.26}_{  -0.28}$&   0.08$^{  +0.34}_{  -0.32}$&  40.42$^{  +0.06}_{  -0.08}$&  13.53$\pm$    0.10&   0.13&   3.80$\pm$   0.70&    1.4&NVSS      &   2.45$^{  +0.04}_{  -0.04}$&  -0.16$^{  +0.02}_{  -0.02}$&  20.80$^{  +0.29}_{  -0.28}$&   1.02$^{  +0.21}_{  -0.18}$\\
RXJ005801.64-275308.6&   5.09$^{  +0.50}_{  -0.52}$&  -0.35$^{  +0.31}_{  -0.29}$&  43.45$^{  +0.04}_{  -0.05}$&  21.43$\pm$    0.15&   0.13&$<$   2.50&    1.4&NVSS      &   1.09$^{  +0.04}_{  -0.04}$&$<$   0.41&  21.55$^{  +0.17}_{  -0.25}$&   3.31$^{  +0.71}_{  -0.54}$\\
RXJ005812.20-274217.8&   0.90$^{  +0.24}_{  -0.22}$&  -0.30$^{  +0.58}_{  -1.02}$&  42.97$^{  +0.10}_{  -0.12}$&  22.63$\pm$    0.30&   0.13&$<$   2.50&    1.4&NVSS      &   1.21$^{  +0.09}_{  -0.09}$&$<$   0.51&  21.75$^{  +0.29}_{  -0.57}$&   0.65$^{  +0.29}_{  -0.17}$\\
RXJ013721.47-182558.3&   3.91$^{  +0.57}_{  -0.46}$&   0.03$^{  +0.19}_{  -0.27}$&  43.22$^{  +0.06}_{  -0.05}$&  20.87$\pm$    0.11&   0.07&$<$   2.50&    1.4&NVSS      &   1.23$^{  +0.04}_{  -0.04}$&$<$   0.36&  20.84$^{  +0.17}_{  -0.15}$&   2.02$^{  +0.30}_{  -0.26}$\\
RXJ014159.22-543037.0&   8.81$^{  +1.51}_{  -1.51}$&  -0.22$^{  +0.41}_{  -0.57}$&  42.98$^{  +0.07}_{  -0.08}$&  17.38$\pm$    0.10&   0.17&$<$  37.00&    5.0&PMN       &   1.60$^{  +0.05}_{  -0.04}$&$<$   0.39&  21.37$^{  +0.20}_{  -0.46}$&   5.98$^{  +1.77}_{  -1.54}$\\
RXJ031456.58-552006.8&   3.45$^{  +0.54}_{  -0.48}$&  -1.66$^{  +0.70}_{  -0.85}$&  43.04$^{  +0.06}_{  -0.06}$&  21.72$\pm$    0.13&   0.18&   1.25$\pm$   0.11&    1.4&FIRST     &   1.10$^{  +0.05}_{  -0.04}$&   0.37$^{  +0.02}_{  -0.02}$&  22.12$^{  +0.10}_{  -0.15}$&   4.42$^{  +1.18}_{  -1.03}$\\
RXJ033340.22-391833.4&   3.37$^{  +0.70}_{  -0.75}$&  -0.24$^{  +0.71}_{  -0.77}$&  44.17$^{  +0.08}_{  -0.11}$&  19.42$\pm$    0.12&   0.10&$<$   2.50&    1.4&NVSS      &   1.52$^{  +0.06}_{  -0.05}$&$<$   0.26&  22.48$^{  +0.19}_{  -0.37}$&   3.01$^{  +1.19}_{  -0.97}$\\
RXJ033402.54-390048.7&   7.22$^{  +0.99}_{  -0.87}$&  -0.06$^{  +0.30}_{  -0.23}$&  42.05$^{  +0.06}_{  -0.06}$&  16.03$\pm$    0.30&   0.10&1567.10$\pm$  52.60&    1.4&NVSS      &   1.84$^{  +0.07}_{  -0.07}$&   0.51$^{  +0.03}_{  -0.03}$&  20.74$^{  +0.29}_{  -0.22}$&   3.97$^{  +0.69}_{  -0.58}$\\
RXJ034119.02-441033.3&   2.60$^{  +0.41}_{  -0.40}$&   0.06$^{  +0.34}_{  -0.34}$&  43.37$^{  +0.06}_{  -0.07}$&  21.55$\pm$    0.11&   0.10&$<$  45.00&    5.0&PMN       &   1.19$^{  +0.04}_{  -0.04}$&$<$   0.72&  21.10$^{  +0.38}_{  -0.31}$&   1.48$^{  +0.34}_{  -0.25}$\\
RXJ043420.48-082136.7&  11.51$^{  +2.69}_{  -1.82}$&  -1.17$^{  +1.21}_{  -0.95}$&  42.97$^{  +0.09}_{  -0.07}$&  18.68$\pm$    0.13&   0.42&   3.70$\pm$   0.60&    1.4&NVSS      &   1.32$^{  +0.05}_{  -0.05}$&   0.24$^{  +0.02}_{  -0.02}$&  21.95$^{  +0.11}_{  -0.22}$&  14.67$^{  +5.80}_{  -4.18}$\\
RXJ045558.99-753229.1&  90.78$^{  +4.93}_{  -3.88}$&  -2.43$^{  +0.26}_{  -0.26}$&  42.08$^{  +0.02}_{  -0.02}$&  16.05$\pm$    0.07&   0.59&$<$  20.00&    5.0&PMN       &   1.34$^{  +0.02}_{  -0.02}$&$<$   0.27&  21.89$^{  +0.04}_{  -0.04}$& 135.47$^{ +13.96}_{ -12.08}$\\
RXJ082640.20+263112.3&   3.81$^{  +0.83}_{  -0.87}$&  -1.27$^{  +0.93}_{  -1.06}$&  42.61$^{  +0.09}_{  -0.11}$&  20.34$\pm$    0.14&   0.25&$<$   0.91&    1.4&FIRST     &   1.27$^{  +0.06}_{  -0.05}$&$<$   0.25&  21.93$^{  +0.15}_{  -0.22}$&   4.57$^{  +2.06}_{  -1.38}$\\
RXJ085340.52+134924.9&  12.00$^{  +0.98}_{  -1.16}$&  -0.19$^{  +0.23}_{  -0.34}$&  43.22$^{  +0.03}_{  -0.04}$&  19.21$\pm$    0.09&   0.25&  23.20$\pm$   1.50&    1.4&NVSS      &   1.26$^{  +0.03}_{  -0.03}$&   0.42$^{  +0.01}_{  -0.01}$&  21.38$^{  +0.15}_{  -0.25}$&   7.80$^{  +1.52}_{  -1.22}$\\
RXJ085851.49+141150.7&   4.05$^{  +0.81}_{  -0.65}$&  -1.32$^{  +0.82}_{  -0.84}$&  43.26$^{  +0.08}_{  -0.08}$&  20.82$\pm$    0.30&   0.29&$<$   2.50&    1.4&NVSS      &   1.20$^{  +0.08}_{  -0.08}$&$<$   0.39&  22.11$^{  +0.13}_{  -0.18}$&   4.81$^{  +1.51}_{  -1.37}$\\
RXJ090518.27+335006.0&   7.58$^{  +2.29}_{  -1.33}$&  -2.43$^{  +1.80}_{  -1.09}$&  43.32$^{  +0.11}_{  -0.08}$&  20.45$\pm$    0.09&   0.16&$<$   0.97&    1.4&FIRST     &   1.17$^{  +0.05}_{  -0.06}$&$<$   0.25&  22.41$^{  +0.08}_{  -0.18}$&  15.65$^{  +6.87}_{  -4.00}$\\
RXJ090923.64+423629.2&   7.40$^{  +0.97}_{  -0.87}$&   0.22$^{  +0.20}_{  -0.22}$&  42.98$^{  +0.05}_{  -0.05}$&  18.25$\pm$    0.09&   0.11&$<$   2.50&    1.4&NVSS      &   1.50$^{  +0.04}_{  -0.03}$&$<$   0.16&  20.69$^{  +0.17}_{  -0.18}$&   3.86$^{  +0.51}_{  -0.50}$\\
RXJ091908.27+745305.6&   4.73$^{  +1.01}_{  -0.92}$&  -0.35$^{  +0.54}_{  -0.82}$&  42.01$^{  +0.08}_{  -0.09}$&  17.82$\pm$    0.09&   0.14&$<$   2.50&    1.4&NVSS      &   1.63$^{  +0.05}_{  -0.05}$&$<$   0.13&  21.57$^{  +0.27}_{  -0.42}$&   4.01$^{  +2.69}_{  -1.61}$\\
RXJ094144.51+385434.8&   2.11$^{  +0.51}_{  -0.53}$&  -0.12$^{  +0.57}_{  -0.58}$&  44.17$^{  +0.09}_{  -0.13}$&  21.11$\pm$    0.14&   0.10&$<$   1.01&    1.4&FIRST     &   1.35$^{  +0.07}_{  -0.06}$&$<$   0.32&  21.92$^{  +0.46}_{  -0.43}$&   1.34$^{  +0.41}_{  -0.37}$\\
RXJ095340.67+074426.1&   4.32$^{  +1.25}_{  -1.09}$&  -0.38$^{  +0.52}_{  -0.94}$&  43.81$^{  +0.11}_{  -0.13}$&  20.01$\pm$    0.09&   0.19&$<$   2.50&    1.4&NVSS      &   1.35$^{  +0.06}_{  -0.06}$&$<$   0.31&  21.77$^{  +0.34}_{  -0.45}$&   3.06$^{  +1.44}_{  -0.95}$\\
RXJ101112.05+554451.3&   6.95$^{  +0.90}_{  -0.88}$&  -0.43$^{  +0.29}_{  -0.29}$&  44.32$^{  +0.05}_{  -0.06}$&  22.02$\pm$    1.00&   0.05& 161.50$\pm$   0.15&    1.4&FIRST     &   1.00$^{  +0.18}_{  -0.17}$&   0.78$^{  +0.07}_{  -0.07}$&  21.62$^{  +0.43}_{  -0.17}$&   4.01$^{  +0.91}_{  -0.57}$\\
RXJ101123.17+524912.4&   3.34$^{  +0.96}_{  -0.87}$&  -1.41$^{  +0.93}_{  -1.22}$&  43.58$^{  +0.11}_{  -0.13}$&  20.86$\pm$    0.09&   0.05&$<$   0.96&    1.4&FIRST     &   1.29$^{  +0.06}_{  -0.06}$&$<$   0.28&  22.51$^{  +0.16}_{  -0.28}$&   3.91$^{  +1.96}_{  -1.21}$\\
RXJ101147.48+505002.2&   4.56$^{  +1.01}_{  -0.82}$&  -0.09$^{  +0.44}_{  -0.47}$&  42.07$^{  +0.09}_{  -0.09}$&  16.74$\pm$    0.22&   0.06&$<$   0.96&    1.4&FIRST     &   1.82$^{  +0.07}_{  -0.07}$&$<$  -0.02&  20.65$^{  +0.73}_{  -0.28}$&   2.75$^{  +0.95}_{  -0.55}$\\
RXJ104723.37+540412.6&   1.71$^{  +0.70}_{  -0.61}$&  -0.56$^{  +0.90}_{  -1.09}$&  43.78$^{  +0.15}_{  -0.19}$&  20.09$\pm$    0.11&   0.06&$<$   0.96&    1.4&FIRST     &   1.54$^{  +0.09}_{  -0.07}$&$<$   0.23&  22.22$^{  +0.42}_{  -0.61}$&   1.31$^{  +0.78}_{  -0.45}$\\
RXJ110742.05+723236.0&   8.11$^{  +1.22}_{  -1.10}$&   0.12$^{  +0.37}_{  -0.45}$&  44.97$^{  +0.06}_{  -0.06}$&  19.07$\pm$    0.08&   0.21& 370.60$\pm$  11.10&    1.4&NVSS      &   1.43$^{  +0.04}_{  -0.04}$&   0.64$^{  +0.01}_{  -0.01}$&  21.58$^{  +0.82}_{  -0.38}$&   4.75$^{  +1.11}_{  -0.75}$\\
RXJ111750.51+075712.8&  16.55$^{  +1.57}_{  -1.40}$&  -0.15$^{  +0.23}_{  -0.29}$&  44.38$^{  +0.04}_{  -0.04}$&  20.67$\pm$    0.09&   0.24&$<$   2.50&    1.4&NVSS      &   1.01$^{  +0.03}_{  -0.03}$&$<$   0.36&  21.57$^{  +0.18}_{  -0.20}$&   9.94$^{  +1.60}_{  -1.26}$\\
RXJ111942.16+211518.1&   3.44$^{  +0.62}_{  -0.48}$&   0.13$^{  +0.30}_{  -0.30}$&  44.24$^{  +0.07}_{  -0.07}$&  19.91$\pm$    0.09&   0.09&$<$   0.93&    1.4&FIRST     &   1.44$^{  +0.04}_{  -0.04}$&$<$   0.21&  21.42$^{  +0.36}_{  -0.28}$&   1.89$^{  +0.37}_{  -0.30}$\\
RXJ121017.25+391822.6&   4.62$^{  +0.68}_{  -0.67}$&  -0.02$^{  +0.35}_{  -0.49}$&  40.98$^{  +0.06}_{  -0.07}$&  17.18$\pm$    0.50&   0.14&$<$   0.98&    1.4&FIRST     &   1.73$^{  +0.10}_{  -0.10}$&$<$   0.04&  21.10$^{  +0.22}_{  -0.49}$&   2.88$^{  +0.78}_{  -0.51}$\\
RXJ121803.82+470854.6&   1.51$^{  +0.44}_{  -0.42}$&  -0.52$^{  +0.60}_{  -1.07}$&  43.82$^{  +0.11}_{  -0.14}$&  21.04$\pm$    0.09&   0.08&$<$   0.98&    1.4&FIRST     &   1.42$^{  +0.07}_{  -0.06}$&$<$   0.31&  22.30$^{  +0.34}_{  -0.69}$&   1.07$^{  +0.53}_{  -0.30}$\\
RXJ124913.86-055906.2&   2.38$^{  +0.47}_{  -0.36}$&  -0.09$^{  +0.57}_{  -0.55}$&  44.36$^{  +0.08}_{  -0.07}$&  17.26$\pm$    0.30&   0.15&$<$   2.50&    1.4&NVSS      &   1.92$^{  +0.07}_{  -0.08}$&$<$   0.12&  22.23$^{  +0.36}_{  -0.57}$&   1.53$^{  +0.39}_{  -0.38}$\\
RXJ131635.62+285942.7&  13.74$^{  +0.98}_{  -1.00}$&   0.35$^{  +0.10}_{  -0.08}$&  43.64$^{  +0.03}_{  -0.03}$&  20.32$\pm$    0.30&   0.08&$<$   0.93&    1.4&FIRST     &   1.10$^{  +0.06}_{  -0.06}$&$<$   0.25&  20.53$^{  +0.07}_{  -0.07}$&   6.55$^{  +0.50}_{  -0.47}$\\
RXJ133152.51+111643.5&  56.99$^{  +4.09}_{  -3.88}$&  -1.72$^{  +0.27}_{  -0.33}$&  43.22$^{  +0.03}_{  -0.03}$&  18.16$\pm$    0.10&   0.13&   8.00$\pm$   1.00&    1.4&NVSS      &   1.17$^{  +0.03}_{  -0.03}$&   0.24$^{  +0.02}_{  -0.02}$&  21.80$^{  +0.06}_{  -0.06}$&  64.44$^{  +9.09}_{  -7.69}$\\
RXJ135105.69+601538.5&   3.83$^{  +0.44}_{  -0.45}$&  -0.12$^{  +0.33}_{  -0.27}$&  43.07$^{  +0.05}_{  -0.05}$&  20.53$\pm$    0.09&   0.12&$<$   2.50&    1.4&NVSS      &   1.27$^{  +0.03}_{  -0.03}$&$<$   0.33&  21.17$^{  +0.23}_{  -0.25}$&   2.22$^{  +0.42}_{  -0.31}$\\
RXJ135529.59+182413.6&   3.63$^{  +0.86}_{  -0.89}$&  -0.41$^{  +0.84}_{  -0.94}$&  44.02$^{  +0.09}_{  -0.12}$&  20.61$\pm$    0.09&   0.14&$<$   2.50&    1.4&NVSS      &   1.31$^{  +0.06}_{  -0.05}$&$<$   0.35&  22.25$^{  +0.25}_{  -0.50}$&   2.82$^{  +1.29}_{  -0.78}$\\
RXJ140134.94+542029.2&   2.94$^{  +0.58}_{  -0.57}$&  -0.13$^{  +0.46}_{  -0.59}$&  41.76$^{  +0.08}_{  -0.09}$&  18.42$\pm$    0.09&   0.08&$<$   1.00&    1.4&FIRST     &   1.63$^{  +0.05}_{  -0.04}$&$<$   0.10&  21.02$^{  +0.39}_{  -0.46}$&   1.89$^{  +0.64}_{  -0.39}$\\
RXJ140416.61+541618.2&   2.79$^{  +0.62}_{  -0.48}$&  -0.08$^{  +0.34}_{  -0.56}$&  44.14$^{  +0.09}_{  -0.08}$&  21.16$\pm$    0.20&   0.08&$<$   0.99&    1.4&FIRST     &   1.28$^{  +0.06}_{  -0.06}$&$<$   0.32&  21.69$^{  +0.40}_{  -0.38}$&   1.70$^{  +0.45}_{  -0.35}$\\
RXJ142754.71+330007.0&   3.00$^{  +0.51}_{  -0.46}$&   0.21$^{  +0.27}_{  -0.39}$&  43.31$^{  +0.07}_{  -0.07}$&  19.66$\pm$    0.08&   0.07&$<$   0.83&    1.4&FIRST     &   1.46$^{  +0.04}_{  -0.04}$&$<$   0.18&  20.81$^{  +0.38}_{  -0.35}$&   1.57$^{  +0.29}_{  -0.27}$\\
RXJ163054.25+781105.1&   8.96$^{  +0.74}_{  -0.87}$&  -0.01$^{  +0.37}_{  -0.29}$&  43.62$^{  +0.03}_{  -0.04}$&  20.54$\pm$    0.14&   0.28&$<$   2.50&    1.4&NVSS      &   1.11$^{  +0.04}_{  -0.04}$&$<$   0.35&  21.52$^{  +0.17}_{  -0.31}$&   5.81$^{  +1.22}_{  -1.06}$\\
RXJ163308.57+570258.7&   1.89$^{  +0.32}_{  -0.33}$&  -0.11$^{  +0.42}_{  -0.69}$&  44.40$^{  +0.07}_{  -0.08}$&  20.06$\pm$    0.12&   0.13&  17.20$\pm$   0.15&    1.4&FIRST     &   1.55$^{  +0.05}_{  -0.04}$&   0.47$^{  +0.01}_{  -0.01}$&  22.48$^{  +0.31}_{  -0.54}$&   1.20$^{  +0.33}_{  -0.26}$\\
RXJ204640.48-363147.5&   6.29$^{  +0.68}_{  -0.71}$&  -0.14$^{  +0.45}_{  -0.36}$&  44.31$^{  +0.04}_{  -0.05}$&  18.76$\pm$    0.07&   0.28&  23.70$\pm$   0.90&    1.4&NVSS      &   1.48$^{  +0.03}_{  -0.03}$&   0.40$^{  +0.01}_{  -0.01}$&  21.89$^{  +0.17}_{  -0.22}$&   3.92$^{  +0.65}_{  -0.57}$\\
RXJ204716.74-364715.1&   9.13$^{  +0.86}_{  -0.91}$&  -0.64$^{  +0.35}_{  -0.53}$&  41.97$^{  +0.04}_{  -0.05}$&  17.57$\pm$    0.08&   0.28&   4.60$\pm$   0.50&    1.4&NVSS      &   1.54$^{  +0.03}_{  -0.03}$&   0.16$^{  +0.01}_{  -0.01}$&  21.53$^{  +0.12}_{  -0.15}$&   7.88$^{  +1.65}_{  -1.37}$\\
RXJ213807.61-423614.3&  16.71$^{  +5.05}_{  -3.62}$&  -1.10$^{  +0.92}_{  -1.06}$&  41.40$^{  +0.11}_{  -0.11}$&  16.15$\pm$    0.07&   0.18&$<$  47.00&    5.0&PMN       &   1.67$^{  +0.05}_{  -0.05}$&$<$   0.32&  21.74$^{  +0.17}_{  -0.27}$&  19.00$^{  +9.30}_{  -6.98}$\\
RXJ235113.89+201347.3&  30.31$^{  +3.16}_{  -3.06}$&  -0.55$^{  +0.29}_{  -0.44}$&  42.37$^{  +0.04}_{  -0.05}$&  16.94$\pm$    0.13&   0.30&   5.90$\pm$   0.50&    1.4&NVSS      &   1.43$^{  +0.04}_{  -0.04}$&   0.14$^{  +0.02}_{  -0.02}$&  21.44$^{  +0.12}_{  -0.18}$&  23.84$^{  +5.35}_{  -4.08}$\\
 &&&&&&&&&&&&\\
 \multicolumn{13}{l}{$^{a}$ Units of erg cm$^{-2}$ s$^{-1}$}\\
 \multicolumn{13}{l}{$^{b}$ See Section \ref{sec:redlums}}\\
 \multicolumn{13}{l}{$^{c}$ Units of erg cm$^{-2}$ s$^{-1}$ keV$^{-1}$}\\
 \end{tabular}
 \end{table*}

\begin{figure}
\begin{center}
\leavevmode
\psfig{figure=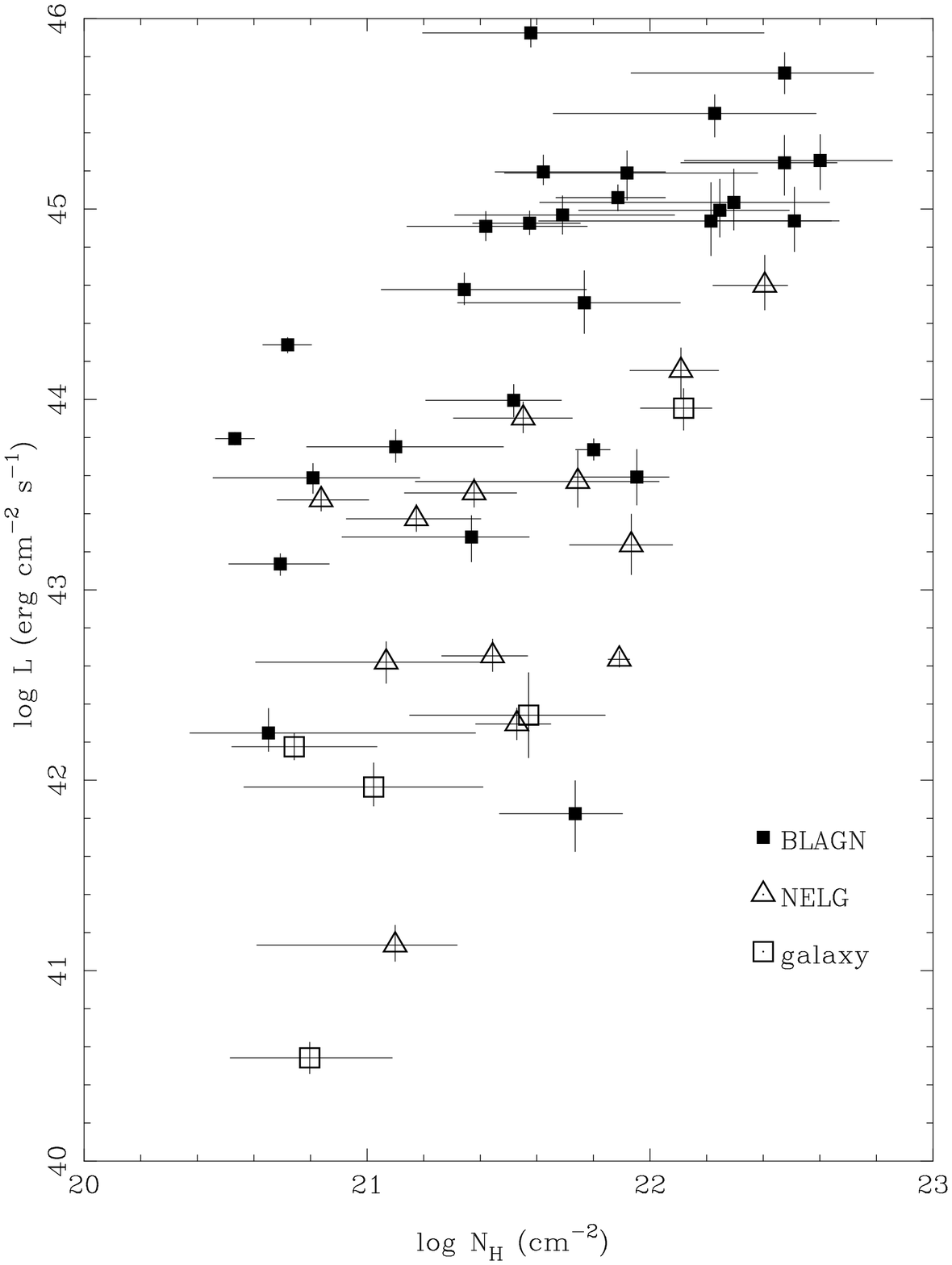,width=80truemm}
\caption{Fitted columns and unabsorbed 0.5-2 keV luminosities for the 
different types of hard sources.}
\label{fig:redcollum}
\end{center}
\end{figure}

The BLAGN, NELG and galaxies' fitted absorbing columns and luminosities (before
absorption) are shown in Fig. \ref{fig:redcollum}.  Most of the high luminosity
sources (near the top of Fig. \ref{fig:redcollum}) are BLAGN, while most of the
galaxies are found with low luminosities; the mean log \(L\) (in erg
s\(^{-1}\), 0.5 - 2 keV) for the BLAGN, NELGs and galaxies are 44.4, 43.2 and
42.2 respectively.
We note that the trend in Fig. \ref{fig:redcollum} for the most absorbed
sources to have the highest luminosity is probably a selection effect due to 
the shifting of the rest frame emitted passband to higher energy with
increasing redshift. This means that sources detectable in the PSPC and
satisfying our spectral selection criterion (see paper 1) will have 
larger columns at higher redshift (and hence at higher luminosity).

\subsection{Notes on individual sources}
\label{sec:notes}

\noindent {\bf RXJ004651.98-204329.0} The hard \ros\ spectrum of this active
galaxy was the subject of Elvis \etal (1997) who proposed that absorption in 
material associated with NGC247 may be responsible.
 
\noindent {\bf RXJ005801.64-275308.6} A second, slightly fainter NELG at J2000
position $00^{h} 58^{m} 1.69^{s} -27^{\rm o} 53' 16.80''$ has the same redshift
($z$= 0.086) as, and is within 10 arcseconds of, the NELG identified as the
X--ray source.

\noindent {\bf RXJ013707.63-183846.4} The star has an unremarkable G type
spectrum, and the X--ray source is coincident with a bright radio source (MRC
0134-188) and hence the star is unlikely to be the correct optical counterpart.
The real optical counterpart is too faint to be seen on POSS.

\noindent {\bf RXJ031456.58-552006.8} This radio galaxy has extended double
lobe structure; see Gruppioni, Mignoli \& Zamorani (1999).

\noindent {\bf RXJ033402.54-390048.7} This source is a wide angle tail radio
source in the galaxy cluster Abell 3135. A magnitude V$\sim$16 Galactic star is
also close to the X-ray source position but is unlikely to be related.

\noindent {\bf RXJ043420.48-082136.7} The identification spectrum of this
object was taken on the WHT as part of RIXOS, but the source was not part of
the final RIXOS sample and is not included in the final catalogue (Mason \etal
2000).

\noindent {\bf RXJ085340.52+134924.9} This RIXOS identification is uncertain
because there is another optical counterpart closer to the centre of the
X--ray error circle that was not observed. Note also that the position of the
optical counterpart is incorrect in Mason \etal (2000), which should be the
same as that in Table \ref{tab:IDs}, \ie $08^{h} 53^{m} 41.01^{s} +13^{\rm o}
49' 19.7''$ (J2000).

\noindent {\bf RXJ095340.67+074426.1} The RIXOS identification and redshift of
this source are based on a relatively poor optical spectrum and therefore may
be incorrect.

\noindent {\bf RXJ101112.05+554451.3} This source was identified as an obscured
radio loud AGN by Barcons \etal (1998). It has strong narrow emission lines but
Mg II $\lambda$2798 is broad, and hence we have classified it as a BLAGN.

\noindent {\bf RXJ101147.48+505002.2} This X--ray source was identified with a
NELG with $z=0.067$ by Carballo \etal (1995), but we identify the source with 
a BLAGN with $z=0.079$, which is brighter and closer to the X--ray
source. Notably, the $z=0.079$ BLAGN showed only narrow lines in the optical
spectrum of Carballo \etal (1995); the \halpha\ line profile appears
genuinely to have changed between 1994 and 1998.

\noindent {\bf RXJ111926.34+210646.1} This galaxy cluster and the target of 
the PSPC observation (PG 1116+215) in which it was found have very similar 
redshifts (0.0176 and 0.1765 respectively) and therefore this source is not
strictly serendipitous.

\noindent {\bf RXJ112056.87+132726.2} The X--ray source is coincident with the
symmetric double radio source 87GB 111822.7+134349, and hence the star
is unlikely to be the correct identification for the X--ray source.

\noindent {\bf RXJ120403.79+280711.2} This galaxy cluster and the target of 
the PSPC observation (PG 1202+281) in which it was found have similar 
redshifts (0.0167 and 0.1653 respectively) and therefore this source may not
be strictly serendipitous.

\noindent {\bf RXJ124913.86-055906.2} This broad absorption line (BAL) QSO was
the only bona-fide BALQSO in sample of Green \& Mathur (1996) to be detected
as a PSPC source.

\noindent {\bf RXJ133146.37+111420.4} This X--ray source is coincident with the
radio source 87GB 132918.1+112918, and hence the star is unlikely to be the
correct optical counterpart.

\noindent {\bf RXJ142754.71+330007.0} The RIXOS identification and redshift of
this source are based on a relatively poor optical spectrum and therefore may
be incorrect.

\section{Discussion}
\label{sec:discussion}

Excluding the clusters, which are unlikely to be an important part of the
faint, hard X--ray source population (see Section \ref{sec:slopeflux}), we have
identified 3 types of extragalactic hard source: BLAGN, NELGs and galaxies. 
At first sight the identification content of the sample appears similar to 
that of other PSPC surveys with similar flux limits but without any spectral
selection, \eg BLAGN being the most numerous source. 
However, in two respects the identification
content of this survey is significantly different.

The first difference is that there is a higher proportion of radio loud objects
(15$^{+7}_{-5}$\%) in this survey (see Section \ref{sec:alphaoxr}).  
There are two possible reasons
why this should be so. One is that the intrinsic X--ray emission of radio loud
sources may be harder than that of radio quiet sources (Reeves \etal 1997) 
because
the radio loud sources have an additional, hard spectrum X--ray emission
component from the inner parts of radio jets. The other possibility is that our
radio loud sources are absorbed, and because they are intrinsically more X--ray
luminous than radio quiet sources they are relatively numerous in a
relatively bright survey of hard spectrum sources. This would be analogous to
the high proportion of (presumably unabsorbed) radio loud sources present in
bright soft X--ray surveys (\eg 11$\pm2$\% in the EMSS, Della Ceca \etal 1994) 
compared to fainter soft
X--ray surveys (\eg 2.5$^{+3.3}_{-1.6}$\% in the CRSS, Ciliegi \etal 1995).

The second difference is that there are nearly half as many NELGs as BLAGN (see
Table \ref{tab:IDs}) compared to ratios of $\sim$ 1:13 NELGS to AGN in RIXOS
(Mason \etal 2000) and $\sim$ 1:6 in the CCRS (Boyle, Wilkes \& Elvis
1997). This implies that the fraction of sources with hard spectra is larger in
the NELG population than in the BLAGN population. This is consistent with the
findings that the NELGs in faint X--ray surveys have, on average, harder X--ray
spectra than BLAGN (Romero-Colmenero \etal 1996, Almaini \etal 1996), and with
the hypothesis that most X--ray selected NELGs are absorbed AGN (Schmidt \etal
1998, Lehmann \etal 2000).

Hence the hypothesis that most of the sources have hard X--ray spectra because
they are absorbed could account for both the high radio loud and NELG content
of this survey. It is also consistent with the preliminary analyses of the
optical spectra and optical colours of the hard sources (Mittaz \etal 2001 and
Carrera \etal 2001).

Related to the absorption hypothesis, an important finding is that the
distributions of X--ray spectral slopes and fluxes of the BLAGN, NELGs and
galaxies are indistinguishable within the current sample, but the three groups
have different ranges of luminosities. The galaxies are found at low luminosity
while the high luminosity sources are BLAGN (see figures \ref{fig:zlumi} and
\ref{fig:redcollum}). The absence of high luminosity, absorbed 
narrow line AGN has also been noted by Akiyama \etal (2000) in the 
\asca\ Large Sky survey.
These results are consistent with all three source types
having the same mechanism for their X--ray spectra (an absorbed active nucleus)
but optical emission line properties which depend on luminosity. 
For this reason we point out that the high redshift, high luminosity, absorbed
QSOs expected to produce a large fraction of the X--ray background are not
neccessarily narrow line objects (the proposed ``QSO 2s'') like Seyfert 2s,
their low luminosity counterparts.

\section{Conclusions}
\label{sec:conclusions}

We have performed a survey of \ros\ fields for serendipitous sources with 
hard spectra
(\(\alpha < 0.5\)); such sources must be a major contributor to the
X--ray background at faint fluxes. In this paper we have presented
optical identifications for 62 of these sources. 
Almost half (28) of these sources are BLAGN, while 12 are NELGs and 5
are galaxies without visible emission lines. 
We have also found 8 clusters of galaxies among the hard
spectrum sources. However, these are predominantly bright sources and are not
particularly hard (their best fit spectral indices all lie in the range \(0.0 <
\alpha < 0.5\)), and hence clusters are unlikely to be an important component
of the hard, faint population.

The hard spectrum BLAGN have a distribution of X--ray to optical ratios which
is similar to that found for AGN from other soft X--ray surveys (\(1 <
\alpha_{OX} < 2\)).  However, a relatively large proportion (15\%) of the
BLAGN, NELGs and galaxies are radio loud. This could be because the radio jets
in these objects produce intrinsically hard X--ray emission, or if they are
absorbed, it could be because radio loud objects are more X--ray luminous than
radio quiet objects.

The BLAGN, NELGs and galaxies have
indistinguishable distributions of X--ray flux and spectra, hence any or all
may be important to the hard, faint population required to solve the XRB
spectral paradox. The majority of the galaxies are low luminosity sources,  
while the highest luminosity objects, and all the high redshift (\(z>1\)) 
sources, are BLAGN.
Their \ros\ spectra are consistent with their being AGN
obscured by columns of \(20.5 < {\rm log} (\nh/{\rm cm^{-2}}) < 23\). 
Overall, our data are consistent with the X--ray emission of the BLAGN, 
NELGs, and the galaxy sources coming from absorbed
active nuclei.

\section{Acknowledgments}

FJC thanks the DGES for partial financial support, under project PB95-0122.
This research was based on observations collected at the European Southern
Observatory, Chile, ESO No. 62.O-0659, and on observations made at the William
Herschel Telescope which is operated on the island of La Palma by the Isaac
Newton Group in the Spanish Observatorio del Roque de los Muchachos of the
Instituto de Astrofisica de Canarias.  This research has made use of data
obtained from the Leicester Database and Archive Service at the Department of
Physics and Astronomy, Leicester University, UK.  This research has made use of
the SIMBAD database, operated at CDS, Strasbourg, France, and of the NASA/IPAC
Extragalactic Database (NED) which is operated by the Jet Propulsion
Laboratory, California Institute of Technology, under contract with the
National Aeronautics and Space Administration.

\section{References}

\refer Akiyama M., \etal, 2000, ApJ, in press (astro-ph/0001289)

\refer Almaini O., Shanks T., Boyle B.J., Griffiths R.E., Roche N., Stewart
G.C., Georgantopoulos I., 1996, MNRAS, 282, 295

\refer Barcons X., Carballo R., Ceballos M.T., Warwick R.S., 
Gonza\/lez-Serrano, 1998, MNRAS, 301, L25

\refer Bohlin R.C., Savage B.D., Drake J.F., 1978, ApJ, 224, 132

\refer Boyle B.J., Wilkes B.J. \& Elvis M., 1997, MNRAS, 285, 511

\refer Carballo R., \etal, 1995, MNRAS, 277, 1312

\refer Carrera F.J., Mittaz J.P.D., Page M.J., 2001, 
proceedings of ``X--ray astronomy '999: Stellar Endpoints, AGN and the
Diffuse Background'', eds G. Malaguti, G. Palumbo \& N. White, in press

\refer Ciliegi P., Elvis M., Wilkes B.J., Boyle B.J., McMahon R.G., Maccacaro
T., 1995, MNRAS, 277, 1463

\refer Condon J.J., Cotton W.D., Greisen E.W., Yin Q.F., Perley R.A., Taylor
G.B., \& Broderick J.J., 1998, AJ, 115, 1693

\refer Della Ceca R., Zamorani G., Maccacaro T., Wolter A., Griffiths R., 
Stocke J.T., Setti G., 1994, ApJ, 430, 533

\refer Elvis M., Fiore F., Giommi P., \& Padovani P., 1997, MNRAS, 291, L49

\refer Fabian A.C., 1999, MNRAS in press

\refer Fabian A.C., Iwasawa K., 1999, MNRAS, 303, L34

\refer Gehrels N., 1986, ApJ, 303, 336

\refer Gilli R., Risaliti G., \& Salvati M., 1999, A\&A, 347, 424

\refer Goodrich R.W., 1989, ApJ, 342, 224

\refer Griffith M.R. \& Wright A.E., 1993, AJ, 105, 1666

\refer Gruppioni C., Mignoli M., \& Zamorani G., 1999, MNRAS, 304, 199

\refer Lehmann I., \etal, 2000, A\&A, in press

\refer Maccacaro T., Gioia I.M., Wolter A., Zamorani G.,  Stocke J.T., 1988,
ApJ, 326, 680

\refer Mason K.O., \etal, 2000, MNRAS, 311, 456

\refer McHardy I.M., \etal, 1998, MNRAS, 295, 641

\refer Mittaz J.P.D., Page M.J., Carrera F.J., 
2001, proceedings of ``X--ray astronomy '999: Stellar Endpoints, AGN and the
Diffuse Background'', eds G. Malaguti, G. Palumbo \& N. White, in press

\refer Mittaz \etal, 1999, MNRAS, 308, 233

\refer Page M.J., Mittaz J.P.D. \& Carrera F.J., 2000, `paper 1', 
MNRAS, 318, 1073

\refer Reeves, J. N., Turner M.J.L., Ohashi T., Kii T., 1997, MNRAS, 292, 468

\refer Romero Colmenero E., Branduardi-Raymont G., Carrera F.J., Jones L.R.,
Mason K.O., McHardy I.M., Mittaz J.P.D., 1996, MNRAS, 282, 94

\refer Schmidt M., Hasinger G., Gunn J., Schneider D., Burg R., Giacconi R.,
Lehmann I., MacKenty J., Tr\" umper J., Zamorani G., 1998, A\&A, 329, 495

\refer Setti G., Woltjer L., 1989, A\&A, 224, L21

\refer White R.L., Becker R.H., Helfand D.J., \& Gregg M.D., 1997, 
ApJ, 475, 479

\refer Wilkes B.J., \etal 1994, ApJS, 92, 53

\refer Zamorani G., \etal 1981, ApJ, 245, 357

\end{document}